\documentclass[runningheads]{llncs} %
\makeatletter

\def\ps@empty{%
  \let\@mkboth\@gobbletwo
  \def\@oddhead{}
  \def\@evenhead{}
  \def\@oddfoot{%
    \parbox[t]{\textwidth}{\centering\scriptsize
    This preprint has not undergone peer review or any post-submission improvements or corrections. The Version of Record of this contribution is published in Computer Safety, Reliability, and Security, SAFECOMP 2026, Lecture Notes in Computer Science, vol 16868, and is available online at https://doi.org/10.1007/978-3-032-34867-8\_20.}}
}
\makeatother

\usepackage[T1]{fontenc}
\usepackage{cite}
\usepackage{orcidlink}
\usepackage{algorithmic}
\usepackage{graphicx}
\usepackage{textcomp}
\usepackage[svgnames]{xcolor}
\usepackage{tcolorbox}
\usepackage{longtable}
\usepackage{listings}
\usepackage{todonotes}
\usepackage{booktabs}
\usepackage{cleveref}
\usepackage{multirow}

\usepackage{bbding}

\newcommand\gitrepourl{\url{https://github.com/embedded-software-laboratory/MISRust}}

\newlength{\ruletablew}
\newlength{\ruletableww}
\begin{document}

\lstdefinelanguage{Rust}{
  morekeywords={
    as, break, const, continue, crate, else, enum, extern, false, fn, for, if, impl, in, let, loop, match, mod, move, mut, pub, ref, return, self, Self, static, struct, super, trait, true, type, unsafe, use, where, while, async, await, dyn
  },
  sensitive=true,
  morecomment=[l]{//},
  morecomment=[s]{/*}{*/},
  morestring=[b]{"},
}

\definecolor{forestgreen}{RGB}{34,139,34}
\definecolor{darkorange}{RGB}{255,140,0}
\definecolor{orangebrown}{RGB}{204, 85, 0}

\lstset{
    language=Rust,
    basicstyle=\ttfamily\footnotesize,
    keywordstyle=\color{blue},
    commentstyle=\color{forestgreen},
    stringstyle=\color{orangebrown},
    string=[b]",
    numbers=left,
    numberstyle=\tiny\color{gray},
    stepnumber=1,
    numbersep=5pt,
    showspaces=false,
    showstringspaces=false,
    tabsize=4,
    breaklines=true,
    breakatwhitespace=false,
    escapeinside={(*@}{@*)}
}

\definecolor{errorcolor}{RGB}{192,57,43}
\definecolor{warningcolor}{RGB}{230,126,34}
\definecolor{infocolor}{RGB}{41,128,185}
\definecolor{codebackground}{RGB}{248,248,248}

\lstdefinestyle{rustcompilererror}{
  backgroundcolor=\color{codebackground},
  basicstyle=\ttfamily\footnotesize,
  breaklines=true,
  captionpos=b,
  commentstyle=\color{infocolor},
  frame=single,
  framesep=2pt,
  keepspaces=true,
  keywordstyle=\color{errorcolor}\bfseries,
  morekeywords={error, warning, note},
  numbers=left,
  numbersep=5pt,
  numberstyle=\tiny\color{gray},
  rulecolor=\color{black},
  showspaces=false,
  showstringspaces=false,
  showtabs=false,
  stringstyle=\color{warningcolor},
  tabsize=2,
  title=\lstname
}

\definecolor{cppkeyword}{RGB}{0, 0, 255}
\definecolor{cppcomment}{RGB}{34, 139, 34}
\definecolor{cppstring}{RGB}{204, 85, 0}
\definecolor{cppnumber}{RGB}{128, 0, 128}
\definecolor{cppbackground}{RGB}{248, 248, 248}

\lstdefinestyle{cppstyle}{
  language=C++,
  backgroundcolor=\color{cppbackground},
  basicstyle=\ttfamily\footnotesize,
  keywordstyle=\color{cppkeyword}\bfseries,
  commentstyle=\color{cppcomment},
  stringstyle=\color{cppstring},
  numberstyle=\tiny\color{gray},
  identifierstyle=\color{black},
  numbers=left,
  numbersep=5pt,
  showspaces=false,
  showstringspaces=false,
  tabsize=4,
  breaklines=true,
  breakatwhitespace=false,
  captionpos=b,
  frame=single,
  framesep=2pt,
  rulecolor=\color{black},
  morekeywords={constexpr, nullptr, noexcept, static_assert, thread_local, int32_t}
}

\definecolor{rustkeyword}{RGB}{0, 0, 255}
\definecolor{rustcomment}{RGB}{34, 139, 34}
\definecolor{ruststring}{RGB}{204, 85, 0}
\definecolor{rustnumber}{RGB}{128, 0, 128}
\definecolor{rustbackground}{RGB}{248, 248, 248}
\definecolor{RedTypename}{RGB}{182,86,17}

\definecolor{rustlifetime}{RGB}{255, 0, 0}
\lstdefinestyle{ruststyle}{
  language=Rust,
  backgroundcolor=\color{rustbackground},
  basicstyle=\ttfamily\footnotesize,
  keywordstyle=\color{rustkeyword}\bfseries,
  commentstyle=\color{rustcomment},
  stringstyle=\color{ruststring},
  numberstyle=\tiny\color{gray},
  identifierstyle=\color{black},
  numbers=left,
  numbersep=5pt,
  showspaces=false,
  showstringspaces=false,
  tabsize=4,
  breaklines=true,
  breakatwhitespace=false,
  captionpos=b,
  frame=single,
  framesep=2pt,
  rulecolor=\color{black},
  ndkeywords={
    bool,u8,u16,u32,u64,u128,i8,i16,i32,i64,i128,char,str,
    Self,Option,Some,None,Result,Ok,Err,String,Box,Vec,Rc,Arc,Cell,RefCell,HashMap,BTreeMap,
    macro_rules
},
ndkeywordstyle=\color{RedTypename},
  morekeywords={as, break, const, continue, crate, else, enum, extern, false, fn, for, if, impl, in, let, loop, match, mod, move, mut, pub, ref, return, self, Self, static, struct, super, trait, true, type, unsafe, use, where, while, async, await, dyn},
  }

\title{MISRust: Mapping MISRA-C++ Coding Guidelines to the Rust Programming Language}
\titlerunning{MISRust: Mapping MISRA-C++ Coding Guidelines to Rust}
\author{Marius Molz\inst{1}\Envelope\orcidID{0009-0005-8406-8727} \and
Niels Schneider\inst{1}\orcidID{0009-0005-5546-4106} \and
Sven Lechner\inst{2}\orcidID{0009-0009-5079-2601} \and
Stefan Kowalewski\inst{1}\orcidID{0000-0001-9397-2009} \and
Alexandru Kampmann\inst{1}\orcidID{0009-0008-8340-1913}}

\authorrunning{M. Molz et al.}
\institute{Chair of Embedded Software, RWTH Aachen University, Aachen, Germany \\
\email{\{molz,schneider,kowalewski,kampmann\}@embedded.rwth-aachen.de}\\
 \and RWTH Aachen University, Aachen, Germany\\
\email{sven.lechner@rwth-aachen.de}\\
}

\maketitle

\begin{abstract}
The Rust programming language is increasingly being considered for safety-critical system development.
However, established safety standards such as ISO 26262 require the use of coding guidelines that do not yet exist for Rust.
This paper systematically examines each of the 179 MISRA C++ 2023 coding guidelines and classifies them into 6 categories based on their applicability to Rust.
Our approach analyzes the rationale behind each MISRA rule to determine whether it remains valid in the Rust programming context.
We find that 47.75\% of the 111 as-is applicable MISRA rules are automatically enforced by Rust's language design, eliminating the need for explicit guideline enforcement.
Furthermore, our analysis explicitly distinguishes between safe and unsafe Rust. We find that 69 guidelines are still relevant and still require either direct application or adaptation for Rust. Importantly, 36 of these rules are automatically satisfied when only using the safe subset of the Rust language. However, they are required again if unsafe Rust features are introduced.
We also identify specific areas where new Rust-specific guidelines are needed. Where a guideline does not directly translate, we propose Rust-specific adaptations that preserve its intent.
All mapping results and supporting artifacts are publicly available as open-source materials at [\gitrepourl].


\end{abstract}

\section{Introduction}
Rust safeguards against entire classes of programming errors through its ownership system, type safety, and memory management model, such as use-after-free, data races, null pointer dereferences, buffer overflows, and uninitialized memory access.
Hence, there is growing interest in using Rust for the development of safety-critical systems.
For example, the Ferrocene project has achieved ISO 26262 tool qualification for a Rust compiler, while automotive initiatives such as Eclipse SDV are incorporating Rust components. \cite{10592287,munch_rust_2025,eclipse.sdv}

Multiple safety standards across domains mandate the use of coding guidelines during software development, 
including ISO 26262 for automotive, DO-178C for aerospace, IEC 62304 for medical devices, 
and EN 50128 for railway systems.
These guidelines limit the usage of specific language features and enforce specific program structures to address typical bugs that occur using a particular language.
Various safety-related coding guidelines exist for C/C++ systems, including MISRA C/C++, AUTOSAR C++ Guidelines, or JSF++ for aerospace. 
However, no equivalent coding guidelines exist for Rust, despite growing interest in safety-critical domains. To date, the MISRA Consortium has only published an addendum to MISRA C:2025 that addresses the applicability of MISRA C:2025 to the Rust programming language, rather than providing a dedicated and comprehensive set of Rust-specific guidelines \cite{misrarust2025,munch_rust_2025}.

The lack of established safety-related coding guidelines is one of the obstacles to achieving compliance with standards when using Rust to develop safety-critical systems. We attempt to address this gap by investigating the transferability of the 179 existing MISRA C++ coding guidelines to Rust. 
We focus on MISRA C++ because it is widely adopted as a coding standard for automotive safety-critical systems and is explicitly recognized in ISO 26262.

Our methodology is based on studying the underlying \textit{rationale} of each MISRA rule that motivates its existence.
Using the rationale, we investigate whether the rule still holds in Rust.
We propose a mapping framework comprising 6 categories to express the degree to which a particular rule is applicable in the Rust context.
Then, for all 179 MISRA rules, we propose a mapping to the 6 categories, ranging from \textit{rule still required in (safe) Rust} to \textit{rule not required, even in unsafe Rust}. Using this methodology, we pursue the following research questions:
\begin{itemize}
    \item \textbf{Which MISRA C++ guidelines remain applicable to Rust?} We provide the first comprehensive mapping of all 179 MISRA C++ 2023 guidelines, identifying 69 guidelines still relevant to Rust. Of them,  58 are directly applicable, while the remaining 11 rules require some form of adaptation for Rust. 
    \item \textbf{What rules can be dropped based on Rust's inherent guarantees?} We demonstrate that 47.75\% of applicable MISRA rules are automatically satisfied by Rust's language design.
    
    \item \textbf{How does unsafe Rust affect guideline compliance?} We identify that 36 guidelines automatically satisfied in safe Rust become unsatisfied in unsafe blocks, providing critical guidance for safety certification of Rust code containing unsafe sections.
\end{itemize}

The remainder of this paper is structured as follows.
Section \ref{sec:related_work} reviews related work.
Section \ref{sec:methodology} presents our methodology and classification framework. 
Section \ref{sec:results_and_analysis} details our mapping results and analysis. 
Section \ref{sec:discussion} discusses implications for safety certification, and Section \ref{sec:conclusion} concludes.

\section{Related Work}
\label{sec:related_work}

The increasing interest in Rust for safety-critical systems development has already led to a few exploratory studies, partial ruleset investigations, and early certification efforts. However, the landscape of coding guidelines regarding Rust in critical domains remains fragmented, with no complete systematic mapping between established safety standards and Rust's unique language design. 

In 2019, Pinho et al. \cite{8990314} provided one of the first works on the applicability of already established safety-focused coding guidelines to Rust. They motivate adopting Rust in safety-critical domains by emphasizing its ownership model, borrow checker, and lifetime system. As part of this investigation, a subset of the MISRA-C guidelines is examined for their relevance to Rust. The study provides analysis results on 9 selected MISRA-C rules. For each rule, the authors explain why it may be simplified when translated to Rust. However, the analysis remains limited in scope and does not attempt to generalize the applicability of rules across the entire rule set. \cite{8990314}

More recently, the MISRA Consortium explicitly addressed the use of Rust in safety-critical contexts. The MISRA C:2025 Addendum 6 ``Applicability of MISRA C:2025 to the Rust Programming Language'' \cite{misrarust2025} provides a high-level overview of how MISRA-C guidelines may translate to Rust. Applicability is evaluated separately for both safe and unsafe Rust code. The addendum also addresses the application of the MISRA C:2025 rule to Rust’s C Foreign Function Interface (FFI), treating it as part of the unsafe category. Nevertheless, as it is only an addendum to the full MISRA-C ruleset, it does not constitute a detailed, rule-by-rule technical mapping, as, e.g., Rust code examples do, nor does it explain the rationale behind the classification decisions. \cite{misrarust2025}

Beyond partial rule mappings, Munch et al. \cite{munch_rust_2025} investigated Rust based on the published MISRA-C Addendum. They analyzed how the Rust programming language differs from C/C++, established safety principles, and identified both enabling factors and open issues. A central challenge highlighted is dynamic memory allocation. The presence of memory allocation, both heap and stack, raises certification concerns, particularly in systems that require predictability. \cite{munch_rust_2025}

In parallel with academic work, industry has begun developing certification-oriented Rust toolchains. Proprietary solutions, such as those from Ferrocene or HighTec, aim to qualify Rust compilers and tools for use in safety-critical system development. However, while Vector and HighTec claim to ``[...] have removed the last hurdles for the use of Rust in the automotive sector''\cite{vector_vh_2024}, it is still unclear how existing established coding guidelines methodically map onto Rust's unique language design. \cite{munch_rust_2025}

\section{Methodology \& Mapping Framework}
\label{sec:methodology}

Rather than evaluating a small subset of rules or focusing solely on conceptual alignment, we aim to provide a broader methodical mapping that supports rigorous reasoning about compliance, applicability, and necessary adaptations when using Rust in safety-critical contexts. We outline the ruleset classification process and discuss the resulting mappings. 

\subsection{Approach}

To ensure sound guideline mappings, we conduct a manual review of each of the 179 MISRA-C++ guidelines. For each guideline, we assess whether its underlying rationale remains applicable in Rust. Our analysis is limited to mapping the MISRA-C++ ruleset to Rust and does not extend to other safety standards. Where a guideline does not directly translate, we propose Rust-specific adaptations that preserve its intent.

If applicable, we present code examples for rules relevant to the mapping decision, which may not be fully compilable or executable. We analyze how the guidelines apply to safe Rust and assess whether these guidelines remain necessary in the context of unsafe Rust. We also propose that runtime panics in Rust can be considered as unsafe behavior. Classification utilizes an analysis template as shown in \Cref{fig:guideline_analysis_template}:

\begin{figure}[htbp]
  \begin{tcolorbox}[colframe=black, colback=white, boxrule=0.5mm, width=\columnwidth]
    \textbf{Rule or Directive and guideline identifier \cite{misracpp2023}}
\begin{longtable}{p{13cm}}
\textbf{Analysis Result} \\
1 of 6 classes that were defined in this work
\end{longtable}

\textbf{Comment:} The comment section explains the rationale behind assigning a specific class in the context of Rust.\\
To support our claims and provide counterexamples where necessary, we may directly include illustrative code examples.
\begin{lstlisting}[style=ruststyle, caption=Rust Code Example for Template Analysis]
  fn main() {
      // This is an example for the guideline analysis template
      println!("Template for guideline analysis");
  }
\end{lstlisting}
\vspace{-3ex}
\end{tcolorbox}
\caption{Template used for guideline analysis}
\label{fig:guideline_analysis_template}
\end{figure}

The fields for the identifier and the specifier (if the guideline is a rule or directive) are taken directly from \cite{misracpp2023}.
The categorization of each guideline into 1 out of the total 6 classes is given in the field \textit{Analysis Result}. Furthermore, the discussion to justify the chosen class is contained in the field named \textit{Comment}, and the code examples are included in the template. 

As far as possible, we provide references from the Rust reference \cite{rust_reference} and the programming handbook \cite{rust_book} for every guideline in the discussion. The specific Rust version being evaluated is \texttt{rustc 1.92.0 (ded5c06cf 2025-12-08)}, edition 2024.

Furthermore, the analysis is confined to Rust language constructs and the standard library. Cross-language interoperability through Rust's FFI is excluded. FFI introduces cross-language interactions that follow C semantics, thereby partially bypassing certain safety guarantees of the Rust compiler and borrow checker. We refer to MISRA C:2025 for an evaluation of the applicability of rules within the context of the Rust FFI.

\subsection{Classification Process}

Our classification procedure is depicted in \Cref{fig:classification-process}. 
We define 6 main classes \textbf{C1 - C6} into which all 179 guidelines can be assigned. The classification begins by determining whether each guideline is applicable to the Rust programming language:

\begin{figure}[htbp]
\centering
\includegraphics[width=0.9\columnwidth]{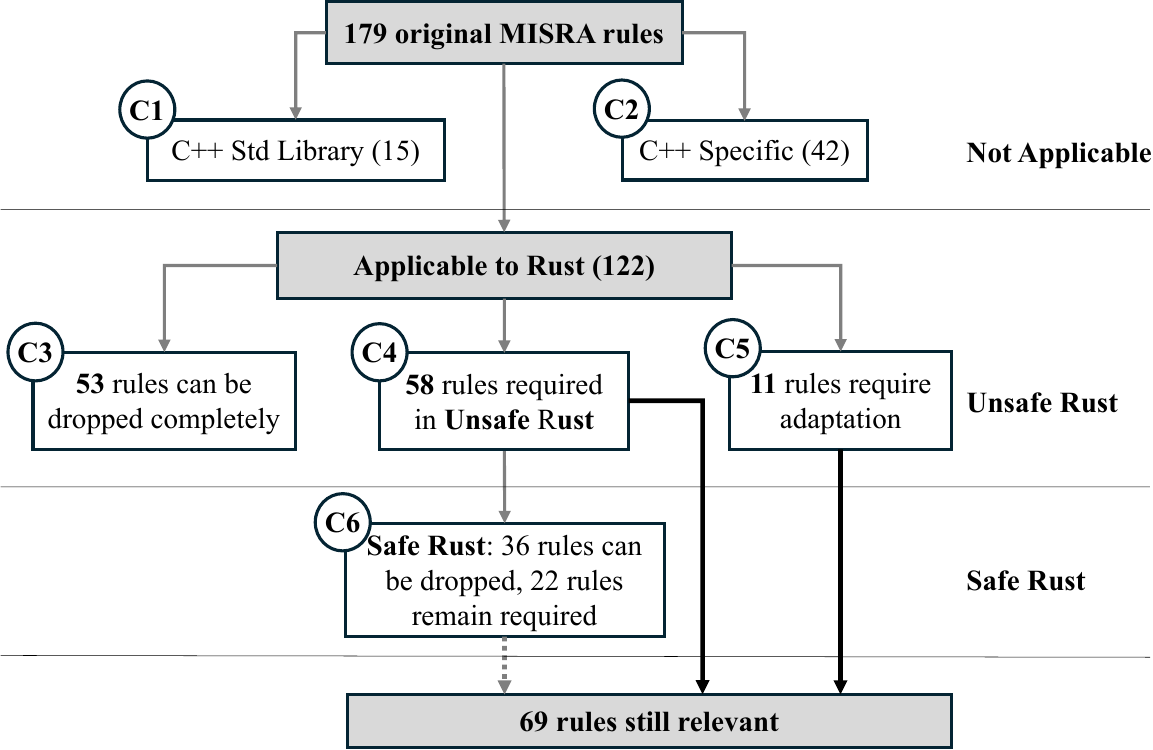}
\caption{Overview of the classification procedure and resulting rule count for each class. Class \textbf{C6} is a strict subset of \textbf{C4}.\label{fig:classification-process}
}
\end{figure}

\begin{enumerate}
    \renewcommand{\theenumi}{(\textbf{C\arabic{enumi}})}
    \item \textbf{Not applicable - C++ Standard Library Usage Restriction:} Guidelines whose rationale is exclusively tied to deficiencies or design issues in the C++ standard library. One such example is  Guideline 25.5.3 ``The pointer returned by the C++ Standard Library functions \texttt{asctime}, \texttt{ctime}, \texttt{gmtime}, \texttt{localtime}, \texttt{localeconv}, \texttt{getenv}, \texttt{setlocale}, or \texttt{strerror} must not be used following a subsequent call to the same function''.

    \item \textbf{Not applicable - C++ Feature does not exist in Rust:} Guidelines that address language features available only in C++. These features are generally considered undesirable or unsafe, and the corresponding guidelines are therefore unnecessary in Rust. In the case of Guideline 12.2.1 ``Bit-fields should not be declared'', there simply does not exist an equivalent feature to \textit{Bitfields} in Rust. Similar missing features in Rust are \texttt{goto}-Statements (Guideline 9.6.1), or virtual pointer-to-member functions (Guideline 13.3.4). \cite{misracpp2023}
\end{enumerate}

After filtering out rules that are not applicable to Rust, we systematically analyze the remaining guidelines to determine their appropriate placement within the final classes \textbf{C3 - C6}. In addition, we identify a dedicated subset, \textbf{C6}, consisting of selected guidelines from \textbf{C4} that require special consideration. We consider these categories to represent the most substantive portion of the ruleset - the guidelines that remain necessary in Rust to preserve the MISRA-C++ ruleset's underlying rationale:

\begin{enumerate}\setcounter{enumi}{2}
    \renewcommand{\theenumi}{(\textbf{C\arabic{enumi}})}
    \item \textbf{Unsafe Rust - Rule can be dropped}: Guidelines in this category are applicable, but can be completely dropped. Rust enforces these Guidelines by design, even in unsafe contexts. One such example is Guideline 19.2.1. The rationale is to prevent confusing behavior and unexpected redefinitions caused by collisions in complex include structures. Due to strict import rules and compile-time checks against redefinition, the problems addressed by this rule are solved at compile time in Rust.

    \item \textbf{Unsafe Rust - Rule still required}: These Guidelines are still required when using unsafe Rust. In general, these are rules that describe broader concepts, such as forbidding recursion or managing dynamic memory. The broadest category \textbf{C4} includes all rules that remain required when writing Rust code that uses unsafe language features. This applies in particular to scenarios where Rust's safety guarantees no longer hold automatically, such as manual memory management, manual pointer interaction, or FFI utilization. Such a rule regarding pointer interaction is Rule 8.2.7, ``A cast should not convert a pointer type to an integral type'': Interpreting a pointer type as an integer yields that pointer's address in Rust. Accessing a pointer itself must be handled in an unsafe block \cite{rust_reference}.

    \item \textbf{Unsafe Rust - Rule requires adaptation}: Guidelines that address C++-specific features that do not exist in Rust, but whose underlying intent would be valuable if expressed in a Rust-specific form. We will present some rule modifications in \Cref{sec:adaptation}.

\item \textbf{Rule still required (safe Rust)}: Guidelines that are still required, even in a safe context. Within category \textbf{C4}, subcategory \textbf{C6} captures the subset of rules that remain necessary even in fully safe Rust. This is especially interesting if one can restrict oneself to only utilizing safe Rust language features. Again, this set also includes rules that address potential programming mistakes at the conceptual level, regardless of the implementation language.

In the case of rule 0.3.1, Rust does not provide suspicious (imprecise) use warnings. Signaling for infinities exists, but the rationale behind the directive is not satisfied. Issues with IEEE 754 floating-point representation are independent of the Rust programming language, and developers must ensure that they prevent unintended software behavior when using Rust.

\end{enumerate}

As shown in \Cref{fig:classification-process}, a strict hierarchy exists among all classification categories. We identify classes \textbf{C4} and \textbf{C5} as the sets of guidelines that remain relevant to preserve the rationale of the MISRA-C++ ruleset within Rust.

\section{Results and Analysis}
\label{sec:results_and_analysis}

In this section, we first give insight into the quantitative results of our classification. Afterwards, we propose possible adaptations for rules that are not applicable as-is, but still required in Rust.

\subsection{Quantitative Overview}

Of the original 179 MISRA rules, we classify 15 as not applicable because they focus solely on compliance with the C++ Standard Library (\textbf{C1}). Additionally, 42 rules that address features available only in C++ are considered not applicable and are classified into class \textbf{C2}.
We therefore identify 122 rules to be applicable to Rust, of which 53 rules are automatically enforced, for example, by Rust's unique ownership model or borrow checker concept. The remaining 69 applicable rules are still required to fulfill the rationale of the MISRA-C++ guidelines. 

\begin{table}[htbp]
    \caption{Distribution of the 179 guidelines across our 6 categories}
    \centering
    \begin{tabular}{@{}lll@{}}
    \toprule
        Category & Guidelines & [\%] \\\midrule
        \textbf{Not applicable:} & 57/179 & 31.84 \\
        \quad (\textbf{C1}) C++ Standard Library Usage Restriction & 15/179 & 08.38 \\
        \quad (\textbf{C2}) C++ Feature does not exist in Rust & 42/179 & 23.46 \\\midrule
        \textbf{Applicable:} & 122/179 & 68.16 \\
        \quad (\textbf{C3}) Rule already satisfied in Rust language & 53/179 & 29.61 \\
        \quad \textbf{Rule still relevant:} & 69/179 & 38.55 \\ 
        \quad \quad (\textbf{C4}) Required in unsafe Rust & 58/179 & 32.40 \\
        \quad \quad \quad (\textbf{C6}) From them, also required in safe Rust & 22/179 & 12.29 \\
        \quad \quad (\textbf{C5}) Rule requires adaptation & 11/179 & 06.15 \\\bottomrule

    \end{tabular}
    \label{tab:guidelines}
\end{table}

As depicted in \Cref{tab:guidelines}, 38.55\% of the 179 MISRA-C++ rules are still relevant. Only 6.15\% of the original ruleset requires adaptation such that the rationale of the rule can be validated in the Rust language (class \textbf{C5}).

Due to the Rust language design, in total, 89 of the 111 as-is applicable rules are automatically satisfied within the safe subset of the language. However, guideline compliance in unsafe Rust, is especially interesting. Of the 89 rules automatically satisfied by safe Rust, 36 rules (\textbf{C4 - C6}) need to be reintroduced in unsafe Rust. 
\begin{table}
    \caption{Distribution of the 69 rules still relevant across the original 179 MISRA-C++ Ruleset. Grouped by topics.}
    \centering
    \begin{tabular}{@{}lll@{}}
    \toprule
        MISRA-C++ topic & Rules still relevant & Topic relevance [\%] \\\midrule
        Language independent issues & 5/10 & 50.00 \\
        General principles & 4/4 & 100.00 \\
        Lexical conventions & 1/12 & 08.30 \\
        Basic concepts & 13/21 & 61.90 \\
        Standard conversions & 5/9 & 55.60 \\ 
        Expressions & 17/24 & 70.80 \\
        Statements & 1/11 & 09.10 \\
        Declarations & 2/8 & 25.00 \\
        Declarators & 3/5 & 60.00 \\
        Classes & 1/4 & 25.00 \\
        Derived classes & 1/6 & 16.70 \\ 
        Member access control & 1/1 & 100.00 \\ 
        Special member functions & 1/8 & 12.50\\ 
        Exception handling & 2/8 & 25.00 \\
        Preprocessing directives & 6/16 & 37.50 \\
        Language support library & 4/12 & 33.30 \\
        Diagnostics library & 1/2 & 50.00 \\
        Algorithms library & 1/5 & 20.00 \\\bottomrule 
    \end{tabular}
    \label{tab:topicdist}
\end{table}

\Cref{tab:topicdist} shows how the 69 rules identified as still relevant (classes \textbf{C3} and \textbf{C4}) are distributed across the original 179 MISRA-C++ Rules, grouped by their respective topic. The distribution reveals highly variant topic relevance. Entire topic sections like these regarding language libraries fall largely into categories \textbf{C1, C2}, which are not applicable to Rust. 
Likewise, object-oriented design topics such as \textit{Classes} (25.00\%), \textit{Derived classes} (16.70\%), and \textit{Special member functions} (12.50\%) are only marginally relevant. This can be attributed to substantial differences in language design, particularly in inheritance and polymorphism. Additionally, Rust does not provide class-based inheritance, classes or class-like structures holding member functions and internal state, or inherited internal variables. 
While Rust allows behavior similar to classes through the use of structs and associated \texttt{impl} blocks, and supports polymorphism via traits acting as interface-like abstractions, these techniques do not permit inheritance of state or class behavior. Especially concepts central to C++, such as accessing inherited members, overwriting base-class functionality, or invoking base-class methods, are not supported in safe Rust. Therefore, many MISRA-C++ Rules addressing class concepts or inheritance are not applicable in Rust. \cite{rust_oop_2025}

\subsection{Unsafe Rust}
\label{sec:unsafe}

While Rust enforces strong guarantees in a safe context, these guarantees are relaxed when using unsafe code. It should be emphasized that the \texttt{unsafe} keyword does not disable borrow checking or all other safety checks, but a few selected relaxations become possible: raw pointer deference, unsafe function calls, interaction with static mutable variables, unsafe trait implementations, and union field access (see \cite{rust_unsafe_2025}). Of the 69 guidelines classified as still relevant to Rust, 36 are violated exclusively in unsafe code contexts.

These violations arise primarily from issues related to lifetimes, access sequencing, and (raw) pointer access. In safe Rust, such properties are statically verified by compiler checks. 

The majority of these affected rules originate from foundational language features such as \textit{Basic concepts}, \textit{Expressions}, or \textit{Standard conversions}. We conclude that unsafe Rust primarily reintroduces risks at a core language level, such as raw pointer arithmetic, and not at a higher level of abstraction. We identify these rule violations to closely resemble classic C or C++ safety issues, such as use-after-free, access sequences, shared resource access, or aliasing violations. This motivates additional research into tools that specifically validate unsafe Rust sections.
    
\subsection{Rules requiring Adaptation}
\label{sec:adaptation}

This category consists of guidelines that require adaptation when applied to Rust. MISRA guidelines \cite{misracpp2023} that fall into this category often refer to features available in Rust that do not map one-to-one to the C++ version. Although the original rationale might still apply, applying the guideline in Rust requires modifications. 

\textbf{Rule 4.1.1: A program shall conform to ISO/IEC 14882:2017(C++17).} At the time of this publication, the Rust programming language does not yet have an ISO/IEC standard. Ongoing efforts working towards a formal specification already exist \cite{rust_reference,rust_issue_spec}. Should such an ISO/IEC standard be created, this rule should be amended to require conformance to it.

\textbf{Rule 6.0.3: The only declarations in the global namespace should be \texttt{main}, namespace declarations and \texttt{extern "C"} declarations.} Part of the rationale for this rule is the influence of include ordering on program behavior. Due to strict rules of unambiguous names when using crates in Rust, this rule is partially satisfied. Prevention of cluttering during lookup still requires following this rule in Rust code (see example in artifacts). A potential adaptation could involve the usage of `prelude' modules that are often used within the Rust ecosystem.

\textbf{Rule 6.4.2: Derived classes shall not conceal functions that are inherited from their bases.} While Rust supports some form of inheritance using traits, Rust does not allow arbitrary method concealing. Traits fundamentally work by exposing all methods that they define on their implementation type. Structs can only shadow trait methods but cannot fully conceal them from consumers. Rust does not allow two functions with the same name and different parameter types. However, in a strict sense, the same behavior is present in Rust as in C++, because traits work the same way in both.

\textbf{Rule 6.5.1: A function or object with external linkage should be introduced in a header file.} The rule rationale indicates that external function or object linkages should be put into a header file and included everywhere to ensure that the underlying type declaration is the same everywhere. The same argument can be extended to Rust. An adaptation of the Rule could involve putting FFI declarations into a dedicated module at the root of a crate named \textit{ffi}. A common approach within the ecosystem involves defining a sister crate with the \textit{-sys} suffix that contains raw FFI bindings, and including the sister crate as a dependency in every project that requires these bindings.

\textbf{Rule 12.3.1: The \texttt{union} keyword shall not be used.} Rust provides \texttt{union} support. However, it suffers from the same drawbacks that the MISRA rule points out. An adaptation of this rule could involve recommending Rust \texttt{enum}s as an alternative to \texttt{union}s. These are type-safe, support \texttt{match} statements, and do not require extensive tracking of \texttt{union} enum variants to prevent undefined behavior.

\textbf{Rule 15.0.1: Special member functions shall be provided appropriately.} Rust facilitates similar semantics using its \texttt{Copy}, \texttt{Clone}, \texttt{Drop}, and \texttt{Pin} traits. An adaptation of a rule should describe the implications of each of these traits in more detail. Additionally, the composition of structs that implement these traits should be noted.

\textbf{Rule 18.5.1: A \texttt{noexcept} function should not attempt to propagate an exception to the calling function.} Rust's error handling does not rely on exceptions. It uses the \texttt{Result} enum to provide type-safe access to error and success variants. However, functions returning a \texttt{Result} are not required to check and handle all error paths within their function body. An adaptation of the rule could require ensuring that all error variants are sufficiently handled within functions or are forwarded to the caller. The adaptation could naturally extend to the usage of panics, restricting them in favor of explicit, recoverable error handling.

\textbf{Rule 19.0.1: A line whose first token is \texttt{\#} shall be a valid preprocessing directive.} While Rust does not use preprocessor directives, it supports conditional inclusion of code via function attributes. As shown in the example file, the rule can be adapted to require that all attributes be valid.

\textbf{Rule 19.2.2: The \texttt{\#include} directive shall be followed by either a \texttt{<filename>} or \texttt{"filename"} sequence.} Because Rust does not use filenames, but rather crate names, and instead of the include directive, the \texttt{use} directive, the guideline is not directly applicable. In general, some restriction on the valid import syntax in Rust is required.

\textbf{Rule 21.6.2: Dynamic memory shall be managed automatically.} While all Rust standard library structs include proper \textit{Drop} trait implementations to free all allocated resources, the rule should be adapted to include guidelines on how to correctly implement the \textit{Drop} for custom structs relying on manual memory allocations.

\textbf{Rule 22.3.1: The \texttt{assert} macro shall not be used with a constant-expression.} While C++ offers a \texttt{static\_assert} macro to execute the assertion at compile-time, Rust's assert macro works in a non-\texttt{const} and \texttt{const} context. This duality can lead to confusion among developers. A potential adaptation is to always wrap assertions intended to be executed within a \texttt{const} context inside a \texttt{const} block, making the intent clear.

\section{Discussion}
\label{sec:discussion}

This section discusses the consequences of adapting our MISRA-C++-motivated Ruleset to Rust with respect to safety certification practice and places our results in context by identifying their validity limitations. We address the observed reduction in applicable required guidelines, the resulting emphasis on governing unsafe code, and the current lack of mature tools and widespread industrial adoption. The discussion also outlines factors that may affect the validity of our analysis, including ongoing language evolution, the subjectivity of manual guideline classification, and the utilization of only a few analysts.

\subsection{Implications for Safety Certification}

Our investigation carries several implications for the use of Rust in safety-critical domains and for establishing a coding guideline similar to the MISRA-C++ guidelines. Most notably, the resulting rule mapping suggests a substantial reduction in the guideline burden. Due to the Rust language design, approximately 61.45\% 
of the original MISRA ruleset are considered unnecessary, as they are automatically fulfilled by the language intrinsics, or not applicable (categories \textbf{C1 - C3}). This strongly indicates that a future Rust-specific safety standard may be significantly more compact, potentially reducing the effort to reach safety compliance.

We identify a critical need for well-defined governance of unsafe Rust features, especially unsafe code blocks. Since \texttt{unsafe} disables some memory safety checks of the Rust compiler, such as raw pointer access \cite{rust_unsafe_2025}, explicit certification guidelines must be defined. Only with such governance extending existing safety concepts can applications written in unsafe Rust be reliably safety-certified.

This paper also identifies a notable safety-related coding guideline deficit in Rust, similar to MISRA-C++. The Rust toolchain already utilizes expressive compiler diagnostics and strict compile-time code checks. Additional tools like the \textit{Clippy} linter claim to further improve code quality \cite{rust_clippy_2025}. However, some MISRA-C++ guidelines have no direct equivalent in the Rust ecosystem. These tools may be a good starting point in extending their static analysis system for future certification frameworks, as new linting rules, compile-time code checks, or even complete third-party analysis tools may be needed to achieve parity with the MISRA-C++ implied safety certification.

While interest in Rust for safety-critical systems is increasing, widespread industrial adoption is still limited \cite{10592287,8990314,eclipse.sdv}. Recent work on using Rust for safety-critical systems reports benefits from the language's integrated memory-safety guarantees, but highlights challenges with legacy code compatibility, the Rust ecosystem's maturity, and the steep learning curve involved in migrating to the new language \cite{munch_rust_2025,8990314,10592287,mueller_leveraging_rust_space_2024}.

In practice, partial integration of Rust and component replacements in a small, limited scope appears to be the most feasible adoption strategy. Small scope changes may allow organizations to leverage Rust whilst reducing risk in the re-certification effort. One prominent example is the introduction of Rust into the Linux Kernel, where it is integrated as a framework for developing device drivers in Rust. The Rust-For-Linux framework extends the existing architecture rather than replacing the C kernel code \cite{10917595}.
Seidel et al. \cite{10592287} also recommend gradual integration of Rust in the context of safety-critical embedded space systems, noting that partial Rust rewrites should be explored in the context of highly critical system components \cite{10592287}.

Institutional research also reflects this incremental adoption pattern. The Eclipse SDV group, for example, has founded a Special Interest Group on Rust, focusing on the adoption in the automotive industry and safety-critical systems \cite{eclipse.sdv}. The DLR Institute of Software Technology has investigated the applicability of Rust in spacecraft systems, highlighting potential benefits and limitations for existing software frameworks, development processes, and standard compliance \cite{sommer2024rust}. 

We therefore conclude that at present Rust is still more likely to complement than substitute established languages, such as C or C++, in safety-critical systems, meaning that widespread adoption is still hindered by a lack of available tooling, standardization, and certification.

\subsection{Threats to Validity}

Many factors may affect the validity of the analysis presented. Both C++ and Rust are evolving languages; future changes in their specifications or common best-practice guidelines may affect how individual MISRA-C++ rules map to similar guidelines for Rust. The classification results presented in this paper only reflect the state of the Rust language at version \texttt{1.92.0}. Therefore, future revisions may be required if new language features emerge.

The classification of the ruleset involved manual interpretation of each individual MISRA-C++ Guideline. Even with our explicit classification template, decisions about whether a guideline is not applicable, automatically satisfied, or needs to be modified to be applicable to Rust (and is still required) depended on the analyst's subjective understanding of the Rust language and its safety concept and semantics. Inconsistent judgments or interpretation bias are therefore possible.

Our study relies on MISRA-C++ as a baseline for deriving an equivalent safety ruleset for Rust. While this provides a sensible starting point, we implicitly assume that the MISRA-C++ design priorities are compatible with the Rust architecture design to a certain level. As the baseline is targeted at C++, and given Rust's unique ownership and borrowing model, some guidelines (classes \textbf{C1 - C3}) are redundant or not applicable per design. 

Finally, the classification was conducted by a small group of analysts \((n=4)\), without independent external verification. This introduces the risk of bias or misclassifications. Future work, therefore, should incorporate a multi-reviewer approach with independent verification to reduce subjectivity and reinforce the robustness of the resulting classifications.

\section{Conclusion}
\label{sec:conclusion}

We have provided a systematic rule-by-rule mapping of all 179 MISRA-C++:2023 guidelines to Rust, introducing a 6-category classification framework to assess each guideline's applicability and necessity in the context of Rust, distinguishing between safe and unsafe Rust. Although safe Rust substantially reduces the guideline burden, unsafe Rust is a critical topic in future safety-focused coding standards.

We identified 11 rules (6.15\% of the original ruleset) that require adaptation to Rust, and proposed possible migrations to preserve the original intention of MISRA-C++.

Overall, we hope that our presented mapping provides a concrete foundation for future standards and tool support, and that it extends the results of the MISRA-C:2025 Addendum. All mapping results and artifacts are made available as open source and accessible at [\gitrepourl].

\bibliographystyle{splncs04}
\bibliography{refs.bib}

\end{document}